\documentclass[lettersize,journal]{IEEEtran}
\usepackage{amsmath,amsfonts}
\usepackage{algorithmic}
\usepackage{algorithm}
\usepackage{array}
\usepackage[caption=false,font=normalsize,labelfont=sf,textfont=sf]{subfig}
\usepackage{textcomp}
\usepackage{stfloats}
\usepackage{url}
\usepackage{verbatim}
\usepackage{graphicx}
\usepackage{cite}
\usepackage{multirow}
\usepackage{booktabs}
\usepackage{xcolor} 
\usepackage{amssymb,bm}
\usepackage{enumitem}

\begin{document}
 
\title{RPRA-ADD: Forgery Trace Enhancement-Driven Audio Deepfake Detection}
\author{Ruibo Fu$^{*}$, \emph{Member, IEEE,}
       Xiaopeng Wang$^{*}$, \emph{Student Member, IEEE, }%
       Zhengqi Wen, \emph{Member, IEEE, }%
       Jianhua Tao, \emph{Senior Member, IEEE, }
       Yuankun Xie, \emph{Student Member, IEEE, }%
       Zhiyong Wang, \emph{Student Member, IEEE, }%
       Chunyu Qiang,\emph{Student Member, IEEE, }%
       Xuefei Liu, \emph{Member, IEEE, }%
       Cunhang Fan, \emph{Member, IEEE, }%
       Chenxing Li, \emph{Member, IEEE, }%
       Guanjun Li, \emph{Member, IEEE}

\thanks{Ruibo Fu, Xuefei Liu, and Guanjun Li are with the Institute of Automation, Chinese Academy of Sciences, Beijing, China. (e-mail: ruibo.fu@nlpr.ia.ac.cn)}
\thanks{Xiaopeng Wang and Zhiyong Wang are with the School of Artificial Intelligence, Chinese Academy of Sciences, Beijing, China.}
\thanks{Yuankun Xie is with the School of Information and Communication Engineering, Communication University of China, Beijing, China.}
\thanks{Zhengqi Wen is with the Beijing National Research Center for Information
Science and Technology, Tsinghua University, Beijing 100084, China.}
\thanks{Jianhua Tao is with the Department of Automation, Tsinghua University, Beijing 100084, China.}
\thanks{Chunyu Qiang is with the School of New Media and Communication, Tianjin University, Tianjin 300072, China}
\thanks{Cunhang Fan is with the Anhui Province Key Laboratory of Multimodal Cognitive Computation, School of Computer Science and Technology, Anhui University, Hefei 230601, China}
\thanks{ Chenxing Li is with the Tencent, AI Lab, Beijing, China}
\thanks{Ruibo Fu and Xiaopeng Wang are co-first authors. Ruibo Fu is the corresponding author. (e-mail: ruibo.fu@nlpr.ia.ac.cn)}}

\markboth{Journal of \LaTeX\ Class Files,~Vol.~XX, No.~XX, December~XXXX}%
{Shell \MakeLowercase{\textit{et al.}}: A Sample Article Using IEEEtran.cls for IEEE Journals}

\maketitle

\begin{abstract}
Existing methods for deepfake audio detection have demonstrated some effectiveness. However, they still face challenges in generalizing to new forgery techniques and evolving attack patterns. This limitation mainly arises because the models rely heavily on the distribution of the training data and fail to learn a decision boundary that captures the essential characteristics of forgeries. Additionally, relying solely on a classification loss makes it difficult to capture the intrinsic differences between real and fake audio.  In this paper, we propose the RPRA-ADD, an integrated Reconstruction-Perception-Reinforcement-Attention networks based forgery trace enhancement-driven robust audio deepfake detection framework. First, we propose a Global-Local Forgery Perception (GLFP) module for enhancing the acoustic perception capacity of forgery traces. To significantly reinforce the feature space distribution differences between real and fake audio, the Multi-stage Dispersed Enhancement Loss (MDEL) is designed, which implements a dispersal strategy in multi-stage feature spaces. Furthermore, in order to enhance feature awareness towards forgery traces, the Fake Trace Focused Attention (FTFA) mechanism is introduced to adjust attention weights dynamically according to the reconstruction discrepancy matrix. Visualization experiments not only demonstrate that FTFA improves attention to voice segments, but also enhance the generalization capability. Experimental results demonstrate that the proposed method achieves state-of-the-art performance on 4 benchmark datasets, including ASVspoof2019, ASVspoof2021, CodecFake, and FakeSound, achieving over 20\% performance improvement. In addition, it outperforms existing methods in rigorous 3×3 cross-domain evaluations across Speech, Sound, and Singing, demonstrating strong generalization capability across diverse audio domains.
\end{abstract}


\section{Introduction}

In recent years, with the rapid development of deepfake technology\cite{vall-E, vits, EELE, F5-TTS}, deepfake audio has become increasingly realistic and harder to distinguish. Deepfake audio technology brings serious challenges to industries like entertainment and news. It has also been misused for fraud, misinformation, and other harmful activities, posing a serious threat to public safety. As a result, audio deepfake detection (ADD) has become an urgent research topic. However, despite the progress achieved by existing ADD methods\cite{oc-softmax}, generalization remains a significant challenge—both within the same audio type (e.g., across different speech datasets) and across different types of audio, such as sound and singing. Existing studies show that the effectiveness of features directly determines the detection capability\cite{feature_1,feature_2,feature_3}. The feature module serves as the foundation of ADD systems, as it captures subtle artifacts introduced by audio forgeries and enables the learning of discriminative representations for reliable detection \cite{survey_chenglong}.

Early efforts focusing on feature design in ADD can be broadly divided into two main categories: (1) Handcrafted Spectral Feature-Based Methods: Traditional approaches often leverage domain knowledge to design acoustic features such as Mel-Frequency Cepstral Coefficients (MFCCs), Linear Frequency Cepstral Coefficients (LFCCs), and Constant-Q Transform (CQT) features \cite{LCNN_LSTM-sum, Attention_Resnet, MCG-RES2Net50, FFT-L-SENet, RawPC, TSSDNet, stable, RawGAT-ST, S2pecNet, AASIST, Graph-ST, DFSincNet, F0, STDC+SE-ResNeXt18}. These features are derived from time-frequency representations of audio signals and offer advantages in computational efficiency and interpretability. However, they rely on manually designed feature selection, and these features are typically designed for specific audio types, making them difficult to adapt to diverse audio content and new forgery techniques \ cite {LFCC-LCNN}. (2) Sinc-Based Temporal Feature Methods: Recent advancements have introduced deep learning models that extract features directly from raw audio waveforms using SincNet-based architectures. Models such as AASIST, RawNet2, and RawGAT-ST \cite{sincnet, AASIST, RawGAT, RawGAT-ST, rawnet2} employ parameterized sinc functions to learn band-pass filters, enabling the capture of discriminative patterns from temporal and spectral cues inherent in audio signals. While these methods provide strong modeling capabilities, they often overfit to specific forgery patterns seen during training, which limits their generalization to unseen attacks. When faced with new forgery patterns outside the training set, both handcrafted and learnable features tend to rely too heavily on surface-level statistical cues, without establishing robust reference points grounded in the intrinsic properties of genuine audio.

To address these limitations, recent advancements in self-supervised learning (SSL), exemplified by models such as Wav2vec and WavLM\cite{HuRawNet2, wav2vec_Large2_XLSR,wavLM,wav2vec2_VIB,wav2vec2-xlsr_ASP_0.31,taslp_xlsr, MAE_AASIST, CodecFake, LLGF,guo2024audio_wavLM,hubert-xl, DARTS, SSAST, WavLM-Large-LSTM, wav2vec2-FC }, have shown promise by learning universal audio representations through large-scale unsupervised pre-training. Unlike traditional handcrafted or task-specific learnable features, SSL-based representations are trained on vast amounts of unlabeled audio data, allowing them to capture general and transferable acoustic patterns. These representations can then be fine-tuned for downstream tasks\cite{wav2ve2_AASIST, wzy_moe_wav2vec2_rawboost}. However, despite their strong representational capacity, existing SSL models are primarily optimized for speech recognition or speaker identification. As a result, their effectiveness in detecting subtle artifacts introduced by audio generation techniques remains limited. Furthermore, directly applying SSL models without adaptation may still lead to suboptimal performance when facing highly dynamic or cross-domain forgery scenarios. This motivates the need to further explore how to enhance the task-awareness and artifact sensitivity of SSL representations in the context of ADD.

In this paper, we propose the RPRA-ADD, an integrated Reconstruction-Perception-Reinforcement-Attention networks based forgery trace enhancement-driven robust audio deepfake detection framework. To enhance the ability to perceive forgery traces, we propose the Global-Local Forgery Perception (GLFP) module. This module extracts global spectral forgery traces through intra-band and inter-band attention and captures local time-frequency features using depthwise separable convolutions, ultimately dynamically merging dual-stream characteristics via a gating mechanism, enabling the capture of subtle forgery traces.
To enhance our method's ability to distinguish the inherent feature differences between real and fake audio, and achieve further feature disentanglement, we design the Multi-stage Dispersed Enhancement Loss (MDEL). In the decoder, MDEL applies a contrastive dispersal strategy to features at different levels, expanding the distribution difference between real and fake audio. Finally, to strengthen this approach's focus on forgery traces, we propose the Fake Trace Focused Attention (FTFA) mechanism, which generates attention weights through reconstruction discrepancy matrices, enhancing the detection of forgery traces. We achieve State-of-the-Art (SOTA) performance on several standard datasets, including ASVspoof 2019, ASVspoof 2021, CodecFake, and FakeSound, achieving over 20\% performance improvement. Furthermore, our method also attains SOTA results in rigorous $3 \times 3$ cross-domain experiments across Speech, Sound, and Singing domains, clearly demonstrating its generalization capability to diverse audio types. The main contributions of this paper are summarized as follows:

\begin{itemize}
\item \textbf{Multi-scale Spatiotemporal Perception}: To uncover key discriminative features containing forgery traces, we propose a dual-stream GLFP module, which uncovers key features containing forgery traces by hierarchically fusing global frequency band information and local temporal information.

\item \textbf{Multi-stage Discrepancy Reinforcement}: To explicitly magnify the forgery differences between real and fake audio during reconstruction, we propose the MDEL, where multi-probe are applied at different stages of feature variation to ensure training stability and accelerate model convergence.

\item \textbf{Forgery Trace Focused Attention}: 
To enhance the focus of features on forgery traces, we introduce the FTFA mechanism, which highlights trace discrepancies by computing their distribution across audio time-frequency patches. This enables explicit localization and targeted enhancement of forged segments, allowing the model to focus on meaningful speech content and reduce interference from silent regions.

\item {\textbf{Cross-Domain Generalization}}: Experimental results demonstrate that the proposed method not only achieves SOTA performance on four mainstream public datasets in in-domain scenarios, with over 20\% performance improvement, but also outperforms existing approaches across diverse audio types (Speech, Sound, and Singing) in rigorous 3×3 cross-domain settings.


\end{itemize}

\section{Related Work}
\subsection{Fake Audio Detection}

ADD mainly focuses on three key categories: speech, sound, and singing. Early methods for Fake Speech Detection (FSpD) relied on handcrafted acoustic features and traditional machine learning algorithms for classification. These methods included manually extracted spectrogram features from raw waveforms \cite{PCDARTS, TSSDNet, RawGAT-ST, AASIST, DFSincNet}, short-term spectral features such as LFCC and MFCC\cite{LFCC, MFCC, FFT-L-SENet, LCNN_LSTM-sum, Attention_Resnet}, long-term spectral features like CQT\cite{CQT-LCNN_21DF_18.31}, subband features such as F0\cite{F0}, and various handcrafted feature fusion approaches\cite{S2pecNet}. However, these handcrafted features are limited by human-designed priors, which often fail to capture the essential generation differences between real and fake audio\cite{spec_feat_shortage}. End-to-end ADD models based on raw waveforms also tend to overfit specific forgery types. Pre-trained models can leverage knowledge from large-scale datasets to improve detection performance. Xie et al. \cite{xie2021siamese_wav2vec} proposed a Siamese network-based model that uses features extracted from the pre-trained Wav2Vec model to distinguish genuine and fake speech. Tak et al.\cite{wav2ve2_AASIST} used Wav2Vec2 in a self-supervised learning form as the frontend and fine-tuned it for ADD tasks. Martin-Donas\cite{martin2022vicomtech_xlsr} utilized deep features from XLS-R, a cross-lingual speech representation model based on Wav2Vec2. Wang et al.\cite{wang23x_interspeech_hubert} extracted prosodic features from HuBERT by modeling speech duration. Guo et al.\cite{guo2024audio_wavLM} adopted wavLM as a frontend feature extractor. However, these pre-trained models—Wav2Vec2, wavLM, and HuBERT—are primarily optimized for speech recognition and tend to focus on semantic information. During pre-training, other important acoustic cues, such as speaker traits or generation artifacts, are often ignored. As a result, using them as feature extractors may introduce error accumulation, as some forgery-relevant information is lost during encoding. Moreover, these models perform poorly on spoofed sounds beyond the speech domain\cite{CodecFake}, highlighting their limited generalization and the need for improved representations in ADD.

Research in Fake Sound Detection (FSoD) has mainly focused on verifying the authenticity of speech. Recently, there has been growing interest in detecting various types of fake sounds. However, FSoD is still an underexplored area. Xie et al. \cite{FakeSound} introduced the FakeSound dataset, which is specifically designed for ADD. It includes a wide range of fake sound scenarios, covering different sound types and Text-to-Audio systems.  In terms of model architecture, Xie et al. used the self-supervised EAT encoder \cite{EAT} and a ResNet-based classification backbone. In the CodecFake dataset \cite{CodecFake}, they generated 49,838 fake samples using the AudioGen model \cite{kreuk2022audiogen}, and constructed a test set with 49,274 real audio clips from AudioCaps \cite{kim2019audiocaps}. They used Wav2Vec2-XLSR \cite{babu22_interspeech_XLSR} for feature extraction and employed AASIST \cite{AASIST} or LCNN \cite{LCNN} as classification backbones.

Fake Singing Detection (FSiD) is a newly emerging subfield within ADD and has received limited attention. FSiD presents challenges that differ from speech-based detection, such as pitch variations constrained by melody, artistic vocal expressions like vibrato and falsetto, and the difficulty of detecting synthetic artifacts in singing with strong instrumental accompaniment or isolated vocals. Xie et al. \cite{FSD} proposed the FSD dataset for detecting fake singing voices in Chinese. Their baseline model combines Wav2Vec2-XLSR with LCNN to identify synthesized singing generated by various synthesis and conversion methods. Later, Zhang et al. \cite{zang2024singfake} introduced the SingFake dataset, which includes both real and fake singing in multiple languages and from diverse singers.

\subsection{Masked Autoencoder}
The Masked Autoencoder (MAE) has emerged as a powerful self-supervised learning framework, demonstrating remarkable performance across both computer vision and speech signal processing domains. Its core principle involves randomly masking portions of input data and leveraging neural networks to reconstruct the missing segments, thereby learning efficient feature representations in an unsupervised manner \cite{he2022masked}. In the image domain, MAE has been extensively applied to tasks such as image classification \cite{image_mae1}\cite{image_mae2}, object detection \cite{object_mae1,object_mae2}, and anomaly detection \cite{anomaly_mae1,anomaly_mae2}, achieving significant advancements. Notably, in deepfake image detection tasks, MAE has demonstrated outstanding performance across multiple benchmark datasets, significantly enhancing the accuracy and robustness of forgery detection \cite{MAE_deepfake1, MAE_deepfake2, MAE_deepfake3 }. 

In the speech domain, MAE has been successfully applied to tasks such as speech enhancement, speech recognition, and speech synthesis. For instance, in speech synthesis, AnCoGen integrates MAE for audio signal reconstruction and transformation, enabling more precise voice conversion \cite{TTS_mae}. In speech enhancement, MAE achieves high-frequency information recovery through spectral masking, improving speech super-resolution \cite{enhance_mae}. In audio classification, MAE combined with Transformer enhances acoustic pattern classification accuracy, making it suitable for environmental sound classification \cite{sound_mae}. In speech emotion recognition, MAE leverages masked audio modeling to extract emotion-related features, enhancing the accuracy and effectiveness of emotion detection \cite{emo_mae}. MAE has achieved satisfactory performance in many domains, but it has yet to be applied in research focused on amplifying forged traces through adversarial reconstruction.

\begin{figure}
    \centering
    \includegraphics[width=1\linewidth]{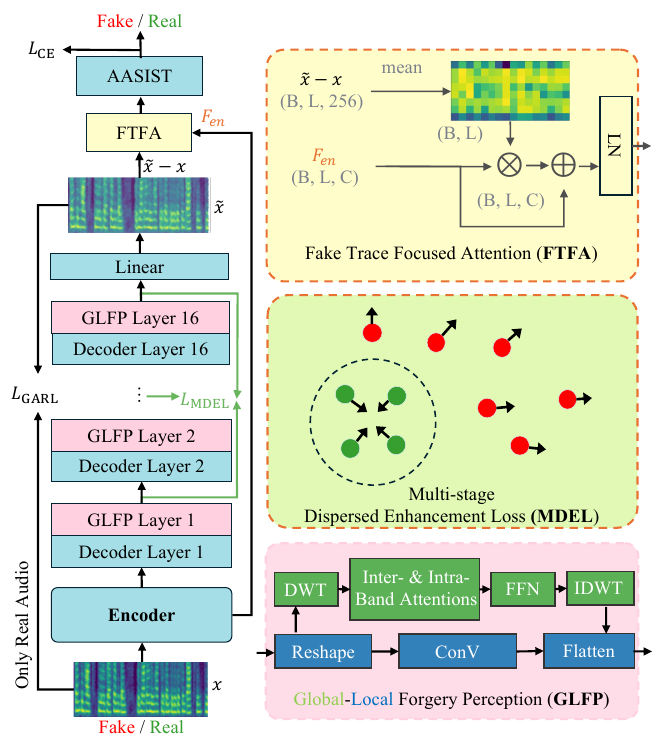}
    \caption{Overall architecture of the proposed RPRA-ADD. The pink part in the figure represents the GLFP module. The green part represents the MDEL. The yellow part indicates the FTFA module.
    }
    \label{fig:REMAE}
\end{figure}

\section{Reconstruction-Perception-Reinforcement-Attention networks for forgery trace enhancement}
Fig.~\ref{fig:REMAE} shows the overall architecture of RPRA-ADD. The model contains a feature encoder, a multi-layer decoder, and the AASIST classifier. It also includes two modules (GLFP and FTFA), and two loss functions (GARL and MDEL). The input is real or fake audio. The encoder extracts features from the audio. These features are sent into a 16-layer decoder. Each decoder layer is followed by a GLFP module. The GLFP module uses discrete wavelet transform (DWT), convolution, and attention (inter-band and intra-band) to detect forgery clues at different scales. During decoding, the Multi-stage Dispersed Enhancement Loss (MDEL) is used. It pushes real samples in the same batch closer and separates real and fake samples in the feature space. This helps the model learn clearer feature differences. At the same time, the Genuine Audio Reconstruction Loss (GARL) is applied. It limits the reconstruction ability to real samples only. This means the model becomes less able to reconstruct fake audio, making it more sensitive to forgery. After decoding, the reconstructed features and the original features are compared. Their difference is sent into the FTFA module. FTFA generates attention weights based on these differences. It highlights local regions with forgery traces. Finally, the enhanced features are passed into the AASIST classifier. The model predicts whether the input audio is real or fake. Cross-entropy loss is used for the final classification.

\subsection{Reconstruction: AudioMAE-Based Encoder-Decoder for Forged simulation reasoning}

To better differentiate between real and fake samples, we use AudioMAE\cite{audioMAE} to reconstruct masked audio patches, aiming to highlight subtle differences.
The input audio is first extracted into fbank features and divided into 16 × 16 patches, resulting in \( N \) patches, where \( N = F \times T \), with \( F \) representing the number of sub-bands and \( T \) the number of temporal segments. Some of these patches are masked, and the set of unmasked patches is denoted as \( X = [x_1, x_2, \ldots, x_N] \in \mathbb{R}^{C \times N} \), where \( C = 256 \) and \( N \) is the number of unmasked patches. The corresponding reconstructed patches are represented as \( \tilde{X} = [\tilde{x}_1, \tilde{x}_2, \ldots, \tilde{x}_N] \in \mathbb{R}^{C \times N} \). Our method leverages the MAE encoder-decoder architecture as the backbone network, where the unmasked patches are input to the encoder, while both the masked patches and the encoded features are passed to the decoder for reconstruction.


Due to the diverse nature of audio forgery methods, it is more effective for the model to focus on the common characteristics of genuine audio rather than overfitting to specific forgery patterns in the training data. To address this, we propose the GARL, which is applied exclusively to genuine speech data. During training, the GARL computes the reconstruction loss for real audio in each mini-batch. The GARL is defined as follows:
\begin{equation}
L_{\text{GARL}} = \frac{1}{N_{\text{R}}} \sum_{i=1}^{N_{\text{R}}} \left( \left( x_{i} - \hat{x}_{i} \right)^2 \right)
\end{equation}
where \( N_{\text{R}} \) represents the number of genuine samples in the mini-batch, \( x_{i} \) denotes the true value of the \(i\)-th genuine sample, and \( \hat{x}_{i} \) is the reconstructed value of the \(i\)-th genuine sample.

\subsection{Perception: Global-Local Forgery Perception Module}
\label{subsec:GLFP}

To strengthen the model's ability to perceive forgery traces, the GLFP module, as shown in Fig.~\ref{fig:REMAE}, consists of two parallel streams: the Global Dynamics Perception Stream and the Local Dynamics Perception Stream. The local stream captures local patterns, while the global stream uses band decomposition and attention mechanisms to extract discriminative micro-artifacts in the frequency domain. Subsequently, a gated fusion mechanism combines complementary features from both streams, enhancing the robustness of the detection process.

\subsubsection{Global Dynamics Perception Stream}
\label{subsubsec:global_stream} 

The global stream extracts frequency artifacts via wavelet decomposition and attention-based feature enhancement. Input features $\mathbf{X}_{in} \in \mathbb{R}^{T \times d}$, conceptually reshaped to a 2D grid ($H \times W$), undergo DWT, decomposing them into low-frequency ($\mathbf{F}_L$) and high-frequency ($\mathbf{F}_{LH}, \mathbf{F}_{HL}, \mathbf{F}_{HH}$) subbands:
\begin{equation}
    \{\mathbf{F}_L, \mathbf{F}_{LH}, \mathbf{F}_{HL}, \mathbf{F}_{HH}\} = \text{DWT}(\text{Reshape}(\mathbf{X}_{in}))
\end{equation}
Following normalization, a dual-attention mechanism processes these subbands: intra-band attention integrates information across bands, while inter-band attention models spatial dependencies within bands. A Feed-Forward Network (FFN) further refines these features, resulting in an enhanced representation $\mathbf{F}_{\text{enhanced}}$. Finally, the global stream output $\mathbf{G}_g$ is reconstructed via Inverse DWT (IDWT):
\begin{equation}
    \mathbf{G}_g = \text{IDWT}(\mathbf{F}_{\text{enhanced}})
\end{equation}
This process amplifies subtle frequency artifacts through attention-based enhancement.

\subsubsection{Local Dynamics Perception Stream}
\label{subsubsec:local_stream} 

The local stream employs a lightweight bottleneck module based on depthwise separable 1D convolutions (DSConv1D) to efficiently model local temporal patterns. This structure significantly reduces computational cost and parameters compared to standard convolutions. Its scaled output, calculated as:
\begin{equation}
    \mathbf{G}_l = s_a \cdot \text{Adapter}_{\text{DSConv1D}}(\mathbf{X}_{in})
\end{equation}
where $\mathbf{X}_{in} \in \mathbb{R}^{T \times d}$ represents the input features and $s_a$ is a fixed scalar, provides fine-grained local context that complements the global stream's frequency-domain analysis.

\subsubsection{Gated Fusion Mechanism}
We dynamically integrate dual-stream features through:

\begin{equation}
    \mathbf{G}_{\text{fused}} = \alpha \cdot \mathbf{G}_g + (1-\alpha) \cdot \mathbf{G}_l
\end{equation}

\begin{equation}
    \alpha = \sigma(\mathbf{W}_g[\mathbf{G}_g; \mathbf{G}_l])
\end{equation}

where $\alpha \in [0,1]$ is the adaptive gating weight, $\mathbf{W}_g$ denotes the fusion matrix, and $\sigma$ is the Sigmoid function. This mechanism strengthens complementary representations in forgery-sensitive regions through feature-aware fusion ratios.

\subsection{Reinforcement: Multi-stage Dispersed Enhancement Loss}

To amplify the forgery traces, we introduce the MDEL to enhance the separation between real and fake samples in the embedding space. This loss encourages real samples to form compact clusters while pushing real and fake samples further apart. To implement this, embeddings from selected transformer layers are concatenated and used to compute pairwise Euclidean distances.

The MDEL consists of two components:
\begin{itemize}
    \item Positive Pair Loss (Real-Real): Minimizes the average Euclidean distance between all pairs of real samples to encourage intra-class compactness.
    \item Negative Pair Loss (Real-Fake): Maximizes the distance between real and fake samples using a hinge-based margin, penalizing those closer than a predefined threshold.
\end{itemize}

The complete loss function is defined as:
\begin{align}
L_{\text{MDEL}} =\ & \frac{1}{N_{RR}} \sum_{\substack{i, j \in R \\ i \ne j}} \|x_i - x_j\|_2 +\notag \\
& \frac{1}{N_{RF}} \sum_{\substack{i \in R \\ j \in F}} \max\left(0, m - \|x_i - x_j\|_2\right)
\end{align}

where \( R \) and \( F \) denote the sets of real and fake samples within each mini-batch, \( N_{RR} \) and \( N_{RF} \) are the numbers of real-real and real-fake sample pairs, respectively, and \( m \) is a hyperparameter controlling the separation margin. This MDEL effectively balances the goals of intra-class compactness and inter-class separation, thereby improving the model’s ability to distinguish genuine speech from forgeries.

By penalizing high similarity between real and fake samples, the MDEL encourages real data to form a tightly clustered distribution in the embedding space while maximizing the disparity between real and fake data. Importantly, no constraints are imposed on the fake data’s compactness, as their features inherently vary due to the diverse nature of forgery methods. This design ensures robustness against a wide range of forgery techniques and enhances the generalization performance of the proposed approach. Since MDEL is computed at each layer output of the MAE decoder, the shallow layers prioritize the reconstruction of important semantic information, while deeper layers focus on the reconstruction of detailed acoustic features.We designed ablation experiments by incorporating MDEL at different decoder layers. The multi-stage loss is computed by aggregating the losses from each layer, and the final loss is obtained by averaging the individual layer losses. This multi-layer approach ensures that both high-level semantic patterns and low-level acoustic details are effectively captured, thus enhancing the overall performance and robustness of the model.

\subsection{Attention: Fake Trace Focused Attention}
To enhance the focus of features on forgery traces, we calculate the absolute difference between the unmasked patches \( X = [x_1, x_2, \ldots, x_N] \) and their corresponding reconstructed patches \( \tilde{X} = [\tilde{x}_1, \tilde{x}_2, \ldots, \tilde{x}_N] \), and use this difference to compute the attention weights. Specifically, we compute the mean absolute difference across all patches to generate the attention weights, normalize these weights using the softmax function, and apply them to the encoded representations. 

The attention weight for the \( i \)-th unmasked patch \( w_i \) is given by:
\begin{equation}
w_i = \text{softmax}\left( \frac{1}{N} \sum_{j=1}^N |x_j - \tilde{x_j}| \right)
\end{equation}
where \( x_j \) and \( \tilde{x_j} \) represent the \( j \)-th unmasked and reconstructed patches, respectively. During training, only unmasked patches are used, and during inference, no masking is applied, so all encoded features \( F_{\text{en}} \) are utilized. The weighted encoded features \( w \cdot F_{\text{en}} \) are then added back to the original encoded features, and the result is passed through a layer normalization step. This operation enhances the model’s focus on important features.

\subsection{Total Loss Function for Co-Training}
The total loss of our framework includes a cross-entropy (CE) loss, GARL, and MDEL. The CE loss is used for final classification, while GARL and MDEL are used to enhance the reconstruction of differences. The total loss \(L_{\text{total}}\) is defined as follows:
\begin{equation}
L_{\text{total}} = L_{\text{CE}} + \lambda_1 L_{\text{GARL}} + \lambda_2 L_{\text{MDEL}}
\end{equation}
where \(\lambda_1\) and \(\lambda_2\) are used to balance the different loss functions. The experiments are shown in Table \ref{table:loss_weights}.

\begin{table*}[t]
    \renewcommand{\arraystretch}{1.5} 
    \caption{Summary of Datasets Used for Evaluation}
    \label{tab:b1}
    \centering
    \begin{tabular}{ccccccc}
    \toprule
    & \textbf{Dataset} & \textbf{Spoofed Type} & \textbf{Unseen Attacks} & \textbf{Spoofed Methods} & \textbf{Real}    & \textbf{Fake}   \\  
    \midrule
    \multirow{2}{*}{Speech} 
        & ASVspoof 19 LA   & TTS, VC         & $\checkmark$                     & 19              & 10256          & 90192  \\
        & ASVspoof 21 LA   & TTS, VC         & $\checkmark$                     & 19              & 14816          & 133360 \\ 
    \midrule
    \multirow{1}{*}{Singing}
        & SingFake         & SVS, SVC        & $\checkmark$                     & Variants of SoVITS 4 & 32312 & 188486 \\ 
    \midrule
    \multirow{2}{*}{Sound}  
        & FakeSound        & TTA, SR         & $\checkmark$                     & 3               & 1899           & 1899   \\
        & CodecFake      & TTA             & $\checkmark$                     & 1               & 49274          & 49838  \\ 
    \bottomrule
    \end{tabular}
    \label{tab:datasets}
\end{table*}

\section{Experimental Setup}
\subsection{Multi-domain Audio Deepfake Datasets }
The evaluation datasets used for the proposed method are detailed in Table \ref{tab:b1}. These datasets include speech, singing, and sound data. The table provides the dataset names, the types of spoofed attacks, the presence of unseen attacks, the number of spoofing methods, and the count of real and fake samples within each dataset.

\subsubsection{\textbf{Speech Deepfake Datasets}} 
For FspD, we train using the ASVspoof 2019 LA (19LA) dataset, a highly influential dataset in FSpD, and use the 19LA development set for validation. To assess the generalizability of the proposed method, we evaluated its performance on various datasets. \textbf{19LA} \cite{todisco2019asvspoof}: This set comprises real and spoofed utterances generated by various TTS and VC algorithms. The test set includes a total of 19 spoofing methods, of which 12 are unseen attacks. ASVspoof 2021 LA (\textbf{21LA} )\cite{yamagishi2021asvspoof}: The 21LA dataset contains utterances transmitted over various channels, enhancing the complexity of the test. This set features a diverse range of state-of-the-art TTS and VC attacks generated using the latest deep learning techniques, designed to test the limits of spoofing countermeasures in realistic scenarios. The test set includes a total of 19 spoofing methods, of which 12 are unseen attacks.

\subsubsection{\textbf{ Sound Deepfake  Datasets}}
     \textbf{FakeSound}\cite{FakeSound}: FakeSound is a dataset containing both real and fake sound data. It uses AudioCaps, a widely used dataset in text-to-audio generation tasks, with sounds synthesized through AudioLDM and SR (Super-Resolution) methods, randomly inserted into real sounds. We use FakeSound for training, validation, and testing, following the training setup proposed in its original method. The test set includes three subsets: Test-Easy, Test-Hard, and Test-Zeroshot. The Test-Easy dataset is set up similarly to the training set, measuring the deepfake detection capability of the proposed method under the same data distribution. The Test-Hard dataset relaxes the constraints on the generated regions, containing audio samples with arbitrary operation durations, varying event lengths, and different levels of generation quality. Test-Zeroshot goes further than Test-Hard, utilizing AudioLDM1, which was not used to generate fake data in the training set.  \textbf{Codecfake} \cite{CodecFake}: Codecfake has a sound dataset generated by the Text to Sound method AudioGen, containing 49,838 fake samples and 49,274 genuine samples. We follow the same experimental setup as Codecfake, using the 19LA dataset for training and validation, and the Codecfake A3 subset for testing.

\subsubsection{\textbf{Singing Deepfake  Datasets}}
    \textbf{SingFake} \cite{zang2024singfake} : SingFake is a multilingual singing dataset. It collects deepfake singing clips from user-generated content websites. The labels indicating whether the singing is real or fake are provided by the uploaders and manually verified by annotators. In this dataset, 60.6\% of the deepfake singing clips are labeled with the “Unknown” generation method, and 92.2\% of those reporting the generation method are variants of SoVITS. The dataset contains 32,312 real samples and 188,486 fake samples.

\subsection{Evaluation Metrics}
In our experiments, we primarily use EER as the evaluation metric. For the ASVspoof 2019 dataset, we additionally report the minimum normalized tandem detection cost function (min t-DCF) \cite{min-DCF}, aligning with other baseline methods. The min t-DCF evaluates the performance of a tandem system consisting of the proposed ADD model and a specific ASV system. For both EER and min t-DCF, lower values indicate better detection performance. In the experiments on the FakeSound dataset, we follow the original protocol and adopt Accuracy (ACC) as the evaluation metric. For cross-domain scenarios, we employ EER to provide a more comprehensive assessment.

\subsection{Implementation Details}
The RPRA-ADD utilizes 128 mel-filterbank features, a window size of 25 ms, and a hop length of 10 ms to convert the waveform into features, corresponding to 10-second audio inputs. Given that prior works, such as \cite{taslp_xlsr} and \cite{wzy_moe_wav2vec2_rawboost}, commonly apply noise enhancement techniques like RawBoost \cite{rawboost} to mitigate noise interference, the proposed method also incorporates different RawBoost approaches for the 21LA datasets to achieve improved performance. The MAE model was initialized with pre-trained parameters from audio-MAE \cite{audioMAE}. AASIST \cite{AASIST} was employed as the back-end classifier. The RPRA-ADD was implemented and trained using the PyTorch framework over 100 epochs. Training was conducted on an NVIDIA RTX 4090 GPU with a batch size of 16 to ensure efficient data processing. The AdamW optimizer was used for optimization, with an initial learning rate of \(5 \times 10^{-6}\). Furthermore, a cosine annealing learning rate decay schedule was applied to dynamically adjust the learning rate throughout the training process, facilitating effective model convergence and improved performance.

\begin{table}[t!]
\centering
 \renewcommand{\arraystretch}{1.5} 
\caption{Performance Comparison of With and Without Pre-Training in 19LA Evaluation}
\label{tab:t3}
\begin{tabular}{ll|cc}
    \toprule
\textbf{System}                 & \textbf{Front-end}   &  \multicolumn{2}{c}{\textbf{19LA}}   \\
                       &              & \textbf{t-DCF}    & \textbf{EER}    \\ \hline
  \textbf{Handcrafted Features}                 \\
LCNN-LSTM \cite{LCNN_LSTM-sum}  & LFCC         & 0.0524    & 1.92              \\
Attention+Resnet \cite{Attention_Resnet} & FFT     & 0.0510    & 1.87          \\
MCG-Res2Net50 \cite{MCG-RES2Net50}  & CQT          & 0.0520    & 1.78           \\
FFT-L-SENet  \cite{FFT-L-SENet}     & FFT          & 0.0368    & 1.14          \\
Graph-ST  \cite{Graph-ST}           & LFB          & 0.0166    & 0.58           \\
F0-SENet34 \cite{FFT-L-SENet}                & F0         & 0.0143    & 0.43          \\ 
STDC+SENet\cite{STDC+SE-ResNeXt18}  & Mel            & -         & 0.22            \\
\hline
  \textbf{End to End}                 \\ 
Stable-AASIST \cite{stable}         &  Waveform &   -       & 1.14           \\
RawGAT-ST    \cite{RawGAT-ST}       &  Waveform & 0.0335    & 1.06           \\
AASIST     \cite{AASIST}            &  Waveform & 0.0275    & 0.83          \\
DFSincNet \cite{DFSincNet}          &  Waveform & 0.0176    & 0.52           \\
  \hline
\textbf{SSL Features} \\
HuRawNet2 \cite{HuRawNet2}                    & Hubert  &0.1393     &1.96  \\
Wav2vec2+AASIST  \cite{wav2ve2_AASIST}        & Wav2vec2  & -         & 0.90           \\
Wav2vec2+LLGF \cite{wav2vec_Large2_XLSR}      & Wav2vec2  & -    &  0.86        \\
Wav2vec2+MoE+AASIST \cite{wzy_moe_wav2vec2_rawboost} & XLSR  & - & 0.74  \\
Wav2vec2+ResNet \cite{taslp_xlsr}             & Wav2vec2  &0.0098 & 0.30  \\ 
GFL-FAD \cite{MAE_AASIST}         & MAE       & 0.0071       & 0.25                    \\    
\hline
\textbf{RPRA-ADD}  & MAE              & \textbf{0.0059}  & \textbf{0.20}      \\

    \bottomrule
\end{tabular}
\label{tab:19_eval}
\end{table}

\section{Experimental Results}
\subsection{In-domain ADD Evaluations}

We conducted training and testing on the \textbf{19LA} speech dataset, which is an authoritative dataset in FSpD commonly used to evaluate the generalization ability of models. In Table \ref{tab:19_eval}, we compare handcrafted features, end-to-end methods, and SSL-based methods. For handcrafted features, we list several recent methods, including LFCC, FFT, LFB, F0, and Mel. Among them, STDC+SENet proposed a spectrum-time fusion method that combines frame-level and utterance-level coefficients, achieving an EER of 0.22\%. For end-to-end methods, DFSincNet introduced an original waveform processing module based on sinc convolutions and multiple pre-emphasis, achieving an EER of 0.52\%. In SSL feature-based methods, we compare SSL models such as Hubert, Wav2Vec2, XLSR, and audioMAE. GFL-FAD achieved an EER of 0.25\% by cascading the audioMAE self-supervised features with the AASIST classification network. Our method also utilizes MAE and AASIST; by incorporating forgery trace enhancement, we further improved performance, achieving an EER of 0.20\% and a min t-DCF of 0.0059.

\begin{table}[h]
\centering
\renewcommand{\arraystretch}{1.5} 
\caption{Comparative EER (\%) results of our proposed method with other fake speech detection systems in the 21LA dataset}
\label{tab:t4}
\begin{tabular}{llc}
    \toprule
    \textbf{Method}       & \textbf{Feature}     & \textbf{EER} (\%)   \\ \hline
    W2V-Large-LLGF \cite{LLGF}      & Wav2Vec2-Large            & 13.19                \\
    WavLM \& MFA \cite{guo2024audio_wavLM}                  & WavLM                     & 10.38                \\ 
    HuBERT-XL \cite{hubert-xl}                      & HuBERT                    & 9.55                 \\
    W2V2-large-DARTS\cite{DARTS}    & Wav2Vec2-Large            & 7.86                 \\
    W2V-XLSR-LLGF  \cite{LLGF}      & Wav2Vec2-XLSR             & 7.26                 \\
    SSAST-CL \cite{ssast_FAD}           & SSAST                     & 4.74                 \\
    WavLM-Large-LSTM~\cite{WavLM-Large-LSTM} & WavLM & 3.50                 \\ 
    Wav2vec + FC~\cite{wav2vec2-FC}          & Wav2Vec2-XLSR & 3.30                \\ 
    Wav2vec2+MoE+AASIST \cite{wzy_moe_wav2vec2_rawboost} & Wav2Vec2-XLSR  & 2.96 \\
    Wav2vec2+ResNet \cite{taslp_xlsr}                    & Wav2Vec2-XLSR  & 2.96 \\ \hline
\textbf{RPRA-ADD}       & MAE          & \textbf{1.19}         \\ \hline
    \bottomrule
\end{tabular}
\end{table}

To better evaluate the generalization capability of our method in the speech domain, we trained our models on the 19LA dataset and tested them on the more challenging \textbf{21LA} dataset. For 21LA, Table \ref{tab:t4} lists the current state-of-the-art self-supervised features, including wav2vec2, WavLM, HuBERT, XLSR, and SSAST, using EER as the evaluation metric. The performance of other comparison methods is directly cited from their respective research papers. Table \ref{tab:t4} reveals that our proposed RPRA-ADD method achieves the best performance compared to the evaluated SOTA self-supervised features on the challenging 21LA dataset.

\begin{table*}[!h]
\centering

\caption{Performance Comparison in SingFake. Integrated System is denoted as IS, Data Augmentation as DA, and External Dataset as ED.Best results are bolded, and second-best results are underlined.}
\label{tab:team_performance}
\renewcommand{\arraystretch}{1.5}

\begin{tabular}{lccc|ccccc|cc|c}
\toprule
\textbf{Method} & \textbf{IS} & \textbf{DA} & \textbf{ED} 
& \textbf{A09} & \textbf{A10} & \textbf{A11} & \textbf{A12} & \textbf{A13} 
& \textbf{Avg EER} \\
\midrule
Fosafer Speech & $\times$ & $\times$ & $\checkmark$ 
& 0.23 & 0.06 & \textbf{0.37} & 4.19 & 0.07 

& \textbf{1.65} \\

NBU\_MISL & $\checkmark$ & $\checkmark$ & $\times$ 
& \textbf{0.13} & 0.11 & 0.94 & 5.17 & 0.10 

& 2.00 \\

I2R-ASTAR & $\checkmark$ & $\checkmark$ & $\times$ 
& 0.65 & 0.51 & 2.49 & 4.57 & 0.64 

& 2.22 \\

Qishan & $\checkmark$ & $\checkmark$ & $\times$ 
& 1.02 & 0.69 & 2.54 & 4.42 & 0.76 

& 2.32 \\

Breast Waves & / & / & / 
& 1.50 & 0.76 & 2.03 & \textbf{4.14} & 0.88 

& 2.73 \\

MediaForensics & / & / & / 
& 0.56 & 0.38 & 3.90 & 4.45 & 1.02 

& 2.75 \\

beyond & / & / & / 
& 0.45 & 0.26 & 4.56 & 4.37 & 0.85 

& 2.99 \\

Star & / & / & / 
& 1.65 & 0.19 & 1.11 & 7.30 & 0.23 

& 3.31 \\
\midrule
RPRA-ADD & $\times$ & $\times$ & $\times$   & \underline{0.17} & \textbf{0.00} & 3.82 & 9.30
& \textbf{0.02}  &4.46\\

\bottomrule
\end{tabular}
\end{table*}

\begin{table}[t]
    \renewcommand{\arraystretch}{1.5} 
    \caption{Performance Comparison on FakeSound Dataset (Accuracy \textuparrow)}
    \label{tab:fakesound}
    \centering
    \begin{tabular}{lccc}
    \toprule
    \textbf{Method} & \textbf{Easy} & \textbf{Hard} & \textbf{Zero-shot} \\ \hline
    Subjective Evaluation       & 0.590 & 0.560 & 0.510 \\
    WavLM \cite{FakeSound}      & 0.790 & 0.580 & 0.610 \\
    EAT \cite{FakeSound}        & \textbf{1.000} & 0.850 & 0.720 \\
    \midrule
    \textbf{RPRA-ADD}           & \textbf{1.000} & \textbf{0.864} & \textbf{0.761} \\
    \bottomrule
    \end{tabular}
\end{table}

\begin{table}[t]
    \renewcommand{\arraystretch}{1.5} 
    \caption{Performance Comparison on CodecFake A3 Task}
    \label{tab:codecfake}
    \centering
    \begin{tabular}{lc}
    \toprule
    \textbf{Method} & \textbf{CodecFake A3 (EER)\textdownarrow} \\ \hline
    Mel-LCNN \cite{CodecFake}     & 31.8  \\
    W2V2-LCNN \cite{CodecFake}    & 49.3  \\
    W2V2-AASIST \cite{CodecFake}  & 45.7  \\
    \midrule
    \textbf{RPRA-ADD}             & \textbf{29.9} \\
    \bottomrule
    \end{tabular}
\end{table}

To evaluate the RPRA-ADD performance in the Sound domain, we conducted training and testing on the \textbf{FakeSound} dataset, with the F1 score as the evaluation metric, as shown in Table \ref{tab:fakesound}. FakeSound includes three test sets: Easy, Hard, and Zeroshot, with increasing difficulty. The human-annotated F1 scores generally range from 0.5 to 0.6, demonstrating that this is a challenging task. We compared the baseline models in the dataset, including wavLM and EAT. Our method achieved SOTA performance on all three test sets, demonstrating the robustness of our approach in the sound domain.

We trained and tested our proposed method on the \textbf{SingFake} singing voice dataset and compared its performance with methods from the SVDD challenge. Some of these methods leveraged additional data, ensemble systems, and data augmentation techniques, while others did not explicitly mention the training strategies used in the SVDD challenge, which we denote as "NA" in the table. In contrast, our method relies solely on a single system, without using data augmentation or additional data, yet still achieved competitive results: SOTA performance on A10 and A13 forgery techniques, and near-optimal performance on A09. 

\subsection{Cross-domain ADD Evaluations}

For the \textbf{CodecFake A3} evaluation,as shown in Table \ref{tab:codecfake}. We regard it as a highly challenging task. Similar to CodecFake \cite{CodecFake}, we utilized the 19LA speech dataset for training without incorporating sound data or applying any data augmentation techniques. Although our method achieved the lowest EER at 29.9\%, 
the relatively high value can be attributed to the substantial domain differences between speech and sound. The distribution of speech data is relatively homogeneous, primarily comprising human vocal recordings captured under specific acoustic conditions. In contrast, sound data presents a much more diverse distribution, including natural sounds, mechanical noises, and music. While real sound data is included in the pre-training dataset, the absence of forged sound data during the training phase limits the model’s ability to adapt to such cases. Consequently, the model faces difficulties in generalizing to the broad variability inherent in sound data, resulting in suboptimal performance.

\begin{table}[h!]
\centering
\renewcommand{\arraystretch}{1.5} 
\caption{Configuration of Baseline Systems}
\label{tab: baseline}
\begin{tabular}{@{}p{2.5cm}p{5cm}@{}}
\toprule
\textbf{Method} & \textbf{Details} \\ \midrule
\textbf{AASIST} & End-to-End Networks \\

\textbf{Wav2Vec2-AASIST} &   SSL+ Classifier    \\
\textbf{audioMAE-AASIST} &   SSL  + Classifier \\ \midrule \addlinespace[0.5em] 
\textbf{RPRA-ADD} &SSL (encoder + decoder) + Classifier \\ \bottomrule
\end{tabular}

\end{table}

In this cross-domain detection performance evaluation, we first considered a variety of baseline architectures, including the end-to-end AASIST and several self-supervised learning (SSL)-based approaches (e.g., XLSR-AASIST, MAE-AASIST, etc.). Detailed configuration information for these models is provided in Table \ref{tab: baseline}. To specifically assess the cross-domain effectiveness of our proposed RPRA-ADD method, we conducted a rigorous comparative experiment. This experiment adopted a domain-specific training and evaluation strategy, in which all models were independently trained on three distinct datasets: Speech (19LA), Environmental Sound (FakeSound), and Singing Voice (SingFake), and were tested within their corresponding domains.

To ensure fair comparisons, all baseline models were retrained using the same training datasets, configurations, and unified hyperparameters as those used for RPRA-ADD. Furthermore, AASIST was employed as the classifier across all models. The results of this comparative evaluation are presented in Table \ref{table:cross_domain_eer}. Experimental findings show that the proposed RPRA-ADD consistently achieves the best overall performance. Regardless of the training domain, RPRA-ADD outperforms all baselines (AASIST, XLSR-AASIST, MAE-AASIST) in terms of AVG EER.

It is also observed that models trained on speech data generally perform poorly when evaluated on the singing voice domain. This may be attributed to the substantial acoustic and stylistic differences between speech and singing data, which hinder the model’s ability to generalize effectively across these domains.

Notably, XLSR-AASIST demonstrates superior performance when trained on singing or environmental sound data and tested on speech. This can likely be attributed to the strong pre-trained speech representations embedded in the XLSR model, which provide a significant advantage in handling speech-related tasks. Nevertheless, despite this specific strength, RPRA-ADD maintains consistent superiority across most in-domain and cross-domain scenarios, as well as in key average performance metrics, thereby demonstrating its overall robustness and superior generalization capability.

\begin{table}[ht]
\centering
\caption{Cross-domain performance of different countermeasures. Each cell reports \textbf{EER (\%)}. Models are trained on a single domain and evaluated on all domains. The last column shows the average across all three domains.}
\label{table:cross_domain_eer}
\begin{tabular}{ll|ccc|c}
\toprule
\textbf{Training} & \textbf{Model} & \textbf{Speech} & \textbf{Singing} & \textbf{Sound} & \textbf{AVG} \\
\midrule
\multirow{4}{*}{Speech} 
 & AASIST              & 1.33   & 45.74   & 49.63   & 32.23 \\
 & XLSR-AASIST         & 1.03   & 44.23   & 26.67   & 23.98 \\
 & MAE-AASIST          & 0.68   & 46.79   & 28.89   & 25.45 \\
 & \textbf{RPRA-ADD}    & \textbf{0.20}  & \textbf{41.90}  & \textbf{21.48}  & \textbf{21.19} \\
\midrule
\multirow{4}{*}{Singing} 
 & AASIST              & 40.79  & 34.25  & 47.41  & 40.82 \\
 & XLSR-AASIST         & \textbf{13.98}  & 10.56  & 44.44  & 22.99 \\
 & MAE-AASIST          & 27.36  & 5.67   & 40.89  & 24.64 \\
 & \textbf{RPRA-ADD}    & 27.11  & \textbf{4.46}   & \textbf{36.67}  & \textbf{22.75} \\

\midrule
\multirow{4}{*}{Sound} 
 & AASIST              & 49.86  & 49.22  & 50.37  & 49.82 \\
 & XLSR-AASIST         & \textbf{43.54}  & 47.09  & 43.70  &44.78  \\
 & MAE-AASIST          & 48.33  & 42.40  & 24.44  & 38.39 \\
 & \textbf{RPRA-ADD}    & 46.70  & \textbf{38.35}  & \textbf{19.26}  & \textbf{34.10} \\
\bottomrule
\end{tabular}
\end{table}

\subsection{Ablation Study}

\subsubsection{Ablation Studies on GLFP Performance}
We conducted ablation studies on the GLFP module, using 19LA as the training set and evaluating on 19LA, SingFake, and FakeSound, as shown in Table \ref{GLFP-ablation}. The results indicate that removing the local branch led to performance degradation across all test sets, particularly on SingFake and FakeSound, highlighting the importance of local modeling for capturing subtle spatiotemporal features of audio forgeries. In contrast, the removal of the global branch primarily affected performance on 19LA, suggesting its role in modeling long-range semantic dependencies. Overall, the complete GLFP module achieved the best performance on all datasets, validating the effectiveness and robustness of the proposed multi-scale dual-stream design in modeling complex forgery patterns.

\subsubsection{MDEL Ablation Experiments}
To determine the optimal configuration for MDEL, we conducted training on the 19LA dataset and performed ablation experiments on the 19LA, SingFake, and FakeSound test sets, as shown in Table \ref{MDEL-ablation}. We evaluated the performance of MDEL under different layer configurations within a 16-layer decoder, including configurations with 0, 1, 2, 4, 8, and 16 layers. A configuration with 0 layers indicates that MDEL was not applied. For the 1, 2, 4, 8, and 16 layer configurations, MDEL was applied as follows: in the 1-layer configuration, DEL was applied only to the last layer; in the 2-layer configuration, DEL was applied to the last two layers; in the 4-layer configuration, DEL was applied every three layers (i.e., at layers 4, 8, 12, and 16); in the 8-layer configuration, DEL was applied every other layer (i.e., at layers 2, 4, 6, 8, 10, 12, 14, and 16); and the 16-layer configuration indicates that DEL was applied to all layers. Experimental results show that MDEL achieves the best performance with the 4-layer configuration. Applying MDEL to more layers introduces unnecessary complexity and does not further enhance the model’s ability to distinguish between real and fake samples.

\begin{table}[ht]
\centering
\caption{Global-Local Ablation Results}
\label{GLFP-ablation}
\setlength{\tabcolsep}{3pt} 
\renewcommand{\arraystretch}{1.2} 
\begin{tabular}{l |c c c | c}
\toprule
\textbf{Config} & \textbf{Speech} & \textbf{Singing} & \textbf{Sound} & \textbf{AVG} \\
\midrule
RPRA-ADD     & 0.20 & 41.90 & 21.48 & 21.19 \\
w/o Global       & 0.41 (\textcolor{red}{+0.21}) & 47.37 (\textcolor{red}{+5.47}) & 22.96 (\textcolor{red}{+1.48}) & 23.58 (\textcolor{red}{+2.39}) \\
w/o Local        & 0.30 (\textcolor{red}{+0.10}) & 48.96 (\textcolor{red}{+7.06}) & 34.82 (\textcolor{red}{+13.34}) & 28.03 (\textcolor{red}{+6.84}) \\
\bottomrule
\end{tabular}
\end{table}

\begin{table}[ht]
\centering
\caption{MDEL Ablation Results}
\label{MDEL-ablation}
\begin{tabular}{c | c c c | c}
\toprule
\textbf{Number of Layers} & \textbf{Speech} & \textbf{ Singing} & \textbf{Sound } & \textbf{AVG} \\
\midrule
0  & 0.63 & 49.93 & 29.63 & 26.73 \\
1  & 0.41 & 45.28 & 25.19 & 23.63 \\
2  & 0.39 & 48.24 & 29.63 & 26.09 \\
4  & \textbf{0.20} & \textbf{41.90} & \textbf{21.48} & \textbf{21.19} \\
8  & 0.38 & 45.60 & 27.41 & 24.46 \\
16 & 0.53 & 51.40 & 31.11 & 27.68 \\
\bottomrule
\end{tabular}
\end{table}

\subsubsection{Multi-stage Dispersed Enhancement Loss for Feature Disentanglement}
\begin{figure}
    \centering
    \includegraphics[width=1\linewidth]{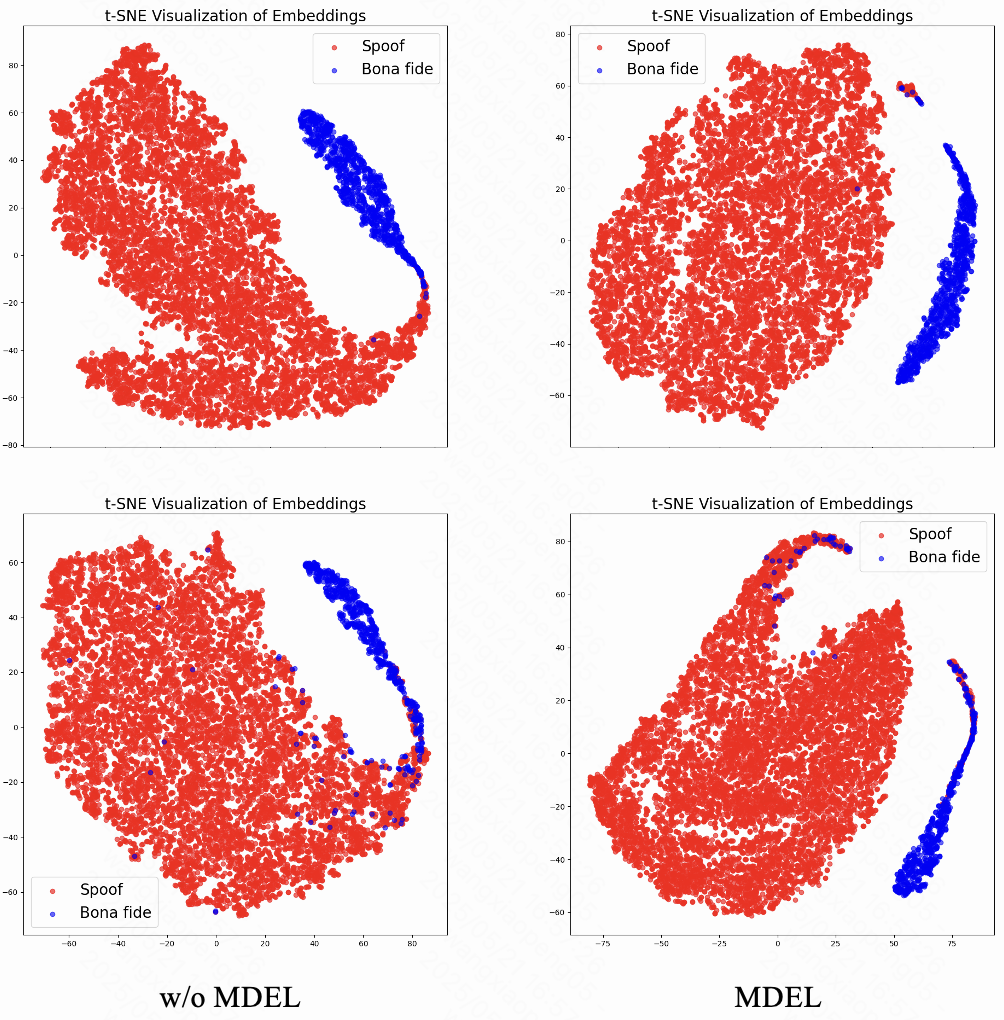}
    \caption{t-SNE visualization of the embeddings with and without MDEL applied. The left side shows the results without MDEL, while the right side shows the results with MDEL. The separation between real and fake samples is more distinct after applying MDEL, and the compactness of the real samples has increased.}
    \label{fig:19-tsne}
\end{figure}

To validate the effect of MDEL in feature disentanglement, we conducted t-SNE visualization of the model embeddings. Specifically, we compared scenarios with and without MDEL on the 19LA and 21LA test sets. t-SNE, a nonlinear technique commonly used for dimensionality reduction in high-dimensional data, allows us to visually inspect the distribution of samples. As shown in Figure \ref{fig:19-tsne}, the left side displays the t-SNE results without MDEL, while the right side shows the results with MDEL applied. Without MDEL, the real and fake samples exhibit no clear separation in the embedding space, with real samples showing a more dispersed distribution, and the distance between the two classes being relatively small, leading to significant overlap. However, after applying MDEL, the compactness of the real samples significantly increases, and the distance between the real and spoof samples grows considerably, demonstrating clear separation. This indicates that MDEL effectively enhances the separation of samples in the feature space, thereby improving the capability of the detection framework to distinguish between real and fake audio.

\begin{table}[h]
\centering
\renewcommand{\arraystretch}{1} 
\caption{Performance Comparison with Different Loss Weights}
\label{table:loss_weights}
\begin{tabular}{ccc|ccc | c}
\toprule
ID & \textbf{ \(\lambda_1\)} &\textbf{  \(\lambda_2\)} & \textbf{Speech} & \textbf{Singing} & \textbf{Sound} & \textbf{AVG}  \\
\midrule
 (a) & 0.1  & 1    & 0.45 & 46.03 & 24.44 & 23.64 \\
 (b) & 0.1  & 0.1  & 0.67 & 46.40 & 33.33 & 26.80 \\
 (c) & 0.1  & 0.01 & 0.55 & 47.36 & 34.07 & 27.33 \\
(d) & 0.01 & 1    & 0.54 & 42.87 & 25.93 & 23.11 \\
(e) & 0.01 & 0.1  & \textbf{0.20} & \textbf{41.90} & \textbf{21.48} & \textbf{21.19} \\
(f) & 0.01 & 0.01 & 0.42 & 47.65 & 25.93 & 24.67 \\

\midrule

\bottomrule
\end{tabular}
\end{table}
\subsubsection{Loss Weights Analysis}
In Table \ref{table:loss_weights}, we explore the hyperparameters of GARL and MDEL. In the experiments, we set \(\lambda_1\) and \(\lambda_2\) to 0.01 and 0.1, respectively. From the experiments, we found that an appropriate \(\lambda_1\) is beneficial, but an excessively high GARL (\(\lambda_1\)=0.1) negatively impacts the results. We speculate that this is because a high GARL reduces the difference between real and fake audio, thus affecting the final classification result. We also conducted experiments on the hyperparameter \(\lambda_2\) of MDEL. When the value of \(\lambda_2\) increased from 0 to 0.01 and 0.1, the performance improved by 8.0\% and 24.0\%, respectively. Overall, the results indicate that careful tuning of \(\lambda_1\) and \(\lambda_2\) is crucial. While a moderate \(\lambda_1\) is beneficial, an excessively high value can be detrimental. Additionally, increasing \(\lambda_2\) from 0 to 0.01 and 0.1 leads to significant performance improvements, highlighting the importance of this parameter in achieving optimal results.

\begin{figure}
    \centering
    \includegraphics[width=1\linewidth]{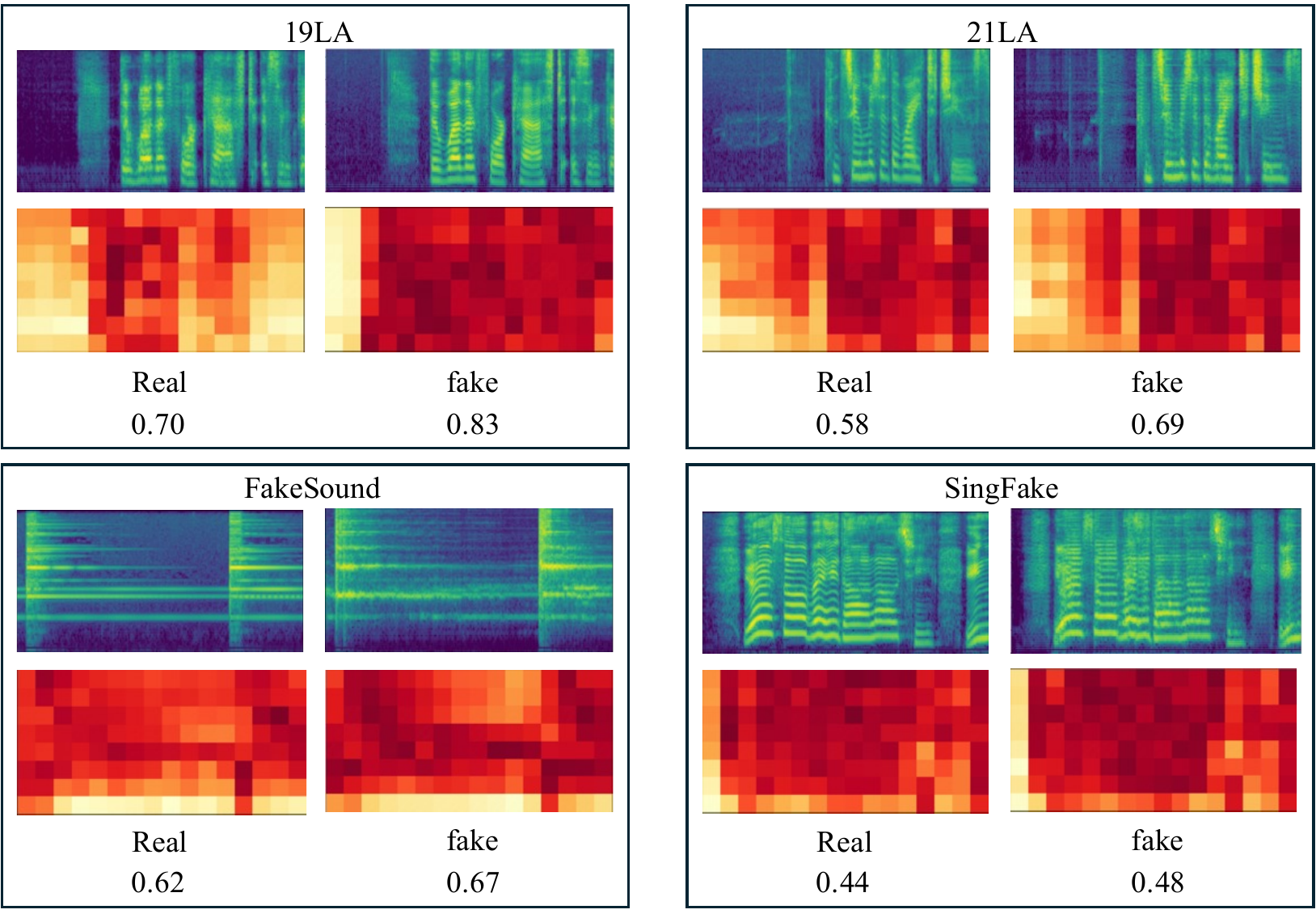}
    \caption{Visualization of FTFA attention weights on the 19LA, 21LA, FakeSound, and SingFake datasets. For each dataset, the top-left image shows the original real fbank feature, and the top-right image shows the fake fbank feature reconstructed by Encodec\cite{encodec}. Below these images, the heatmap shows the difference between the original and reconstructed features, as calculated by the FTFA module. Brighter red colors mean higher attention weights, which means the model focuses more on those patches. The numbers below each heatmap are the average difference values between the original and reconstructed fbank features.}
    \label{fig:weights}
    
\end{figure}

\begin{figure}
    \centering
    \includegraphics[width=1\linewidth]{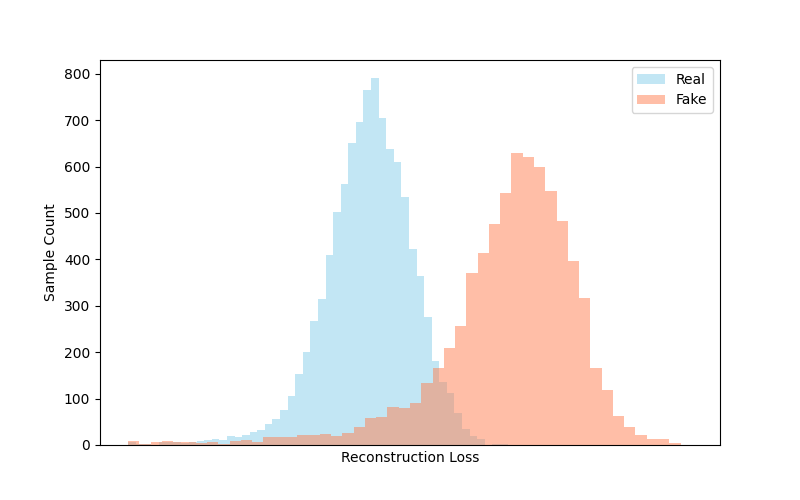}
    \caption{Distribution of reconstruction losses for real and fake samples computed using the RPRA-ADD method.}
    \label{fig:loss_recon}
\end{figure}
\subsection{Visualization of Discriminative Feature Representations}
We systematically visualized the reconstructed patch-level attention matrices on the 19LA, 21LA, FakeSound, and SingFake datasets. In each subfigure, the left side shows the fbank features and corresponding difference matrix of real samples, while the right side shows those of fake samples. The numbers below represent the average difference between the original and reconstructed fbank features (mean reconstruction error). As shown in Fig.\ref{fig:weights}, the attention weights are clearly focused on regions with higher energy. In the 19LA and 21LA (speech data) and SingFake (singing data) datasets, the model pays more attention to time frames containing sound activity, while the weights on silent parts are significantly lower. This agrees with the prior knowledge of speech forgery detection, where the focus should be on sound regions rather than silent segments. Previous studies \cite{Silence_FAD} have shown that removing silent parts can significantly reduce the performance of some models. The introduction of the FTFA module effectively alleviates this problem and improves the model’s sensitivity to the complete audio stream.

In Figure \ref{fig:loss_recon}, we present the distribution of reconstruction losses for real and fake samples computed by the RPRA-ADD method. The horizontal axis represents the reconstruction difference, while the vertical axis indicates the number of samples. It can be observed that the reconstruction losses for real samples (shown in blue) are mainly concentrated in the lower range, whereas those for fake samples (shown in red) are relatively higher. This clear difference in reconstruction losses enables our model to effectively distinguish between real and fake samples.

\section{Conclusion}
In this work, we propose RPRA-ADD, a novel forgery trace enhancement-driven framework for robust audio deepfake detection. By integrating GLFP, MDEL, and FTFA, the RPRA-ADD effectively captures subtle forgery artifacts while improving feature discrimination.  Experiments demonstrate that RPRA-ADD achieves SOTA performance on four benchmarks and exhibits strong generalization across diverse audio types and domains. We conducted visualization experiments to further analyze the proposed method, which demonstrates that the model effectively attends to critical speech segments containing forgery traces, thereby validating its ability to localize and enhance discriminative. In the future, we will continue to explore trace enhancement techniques in multimodal deepfake detection methods, focusing on real-world data.

\bibliography{main}
\bibliographystyle{IEEEtran}

\end{document}